\PassOptionsToPackage{table}{xcolor}
\documentclass[acmtog]{acmart}

\usepackage{graphicx} 
\usepackage{amsmath}
\usepackage{enumitem}
\usepackage{multirow}
\usepackage{algorithm}
\usepackage{algpseudocode}

\newcommand{\ignore}[1]{}

\definecolor{green}{rgb}{0, 0.5, 0}
\definecolor{orange}{rgb}{0.6, 0.3, 0.1}
\definecolor{red}{rgb}{1.0, 0.0, 0.0}
\definecolor{teal}{rgb}{0.0, 0.4, 0.4}
\definecolor{purple}{rgb}{0.65,0,0.65}
\definecolor{saffron}{rgb}{0.8,0.55,0.1}
\definecolor{turquoise}{rgb}{0.0,0.5,0.5}
\definecolor{brown}{rgb}{0.5, 0.16, 0.16}
\definecolor{brickred}{rgb}{.6, .2 .1}
\definecolor{coral}{rgb}{1,0.45,0.33}

\definecolor{rankonecolor}{HTML}{FFDADA}   
\definecolor{ranktwocolor}{HTML}{CAEEFB}   

\newcommand{\rankone}{\cellcolor{rankonecolor}}
\newcommand{\ranktwo}{\cellcolor{ranktwocolor}}

\copyrightyear{2026}
\acmYear{2026}
\setcopyright{cc}
\setcctype{by}
\acmConference[SA Conference Papers '26]{SIGGRAPH Asia 2026 Conference Papers}{December 01--04, 2026}{Kuala Lumpur, Malaysia}
\acmBooktitle{SIGGRAPH Asia 2026 Conference Papers (SA Conference Papers '26), December 01--04, 2026, Kuala Lumpur, Malaysia}
\acmDOI{10.1145/3829340.3842246}
\acmISBN{979-8-4007-2842-6/2026/12}

\begin{document}

\title{Architectural Textured Proxy Generation}

\author{Weidan Xiong}
\affiliation{
 \department{Guangdong Provincial Key Laboratory of Visual Media and Multidimensional Intelligence, College of Computer Science and Software Engineering}
	\institution{Shenzhen University}
	\country{China}	
}
\email{xiongweidan@gmail.com}

\author{Shihao Feng}
\affiliation{
	\institution{Shenzhen University}
	\country{China}	
}
\email{shihao.feng.vcc@gmail.com}

\author{Kaijie Qu}
\affiliation{
	\institution{Shenzhen University}
	\country{China}	
}

\author{Hao Li}
\affiliation{
	\institution{Shenzhen University}
	\country{China}	
}

\author{Mingqi Zhang}
\affiliation{
	\institution{Shenzhen University}
	\country{China}	
}

\author{Daniel Cohen-Or}
\affiliation{
	\institution{Shenzhen University}
	\country{China}	
}

\author{Hui Huang}
\authornote{Corresponding author: Hui Huang (hhzhiyan@gmail.com)}
\affiliation{%
     \department{Guangdong Provincial Key Laboratory of Visual Media and Multidimensional Intelligence, College of Computer Science and Software Engineering}
	\institution{Shenzhen University}
	\country{China}}
     \email{hhzhiyan@gmail.com}

\begin{abstract}
Textured proxy models are a fundamental representation for architectural scenes, enabling efficient rendering while preserving visual fidelity. Existing approaches construct such proxies through a sequential pipeline in which geometry is first abstracted from a dense reconstruction and textures are later projected from input images. This geometry-only formulation neglects image evidence during proxy generation, leading to misalignment of structural features such as facades, edges, and their intersections across views. As a result, inconsistencies are deferred to a downstream correction stage during texturing, where the underlying geometric structure can no longer be reliably recovered.
In this work, we introduce an image-guided proxy formulation in which proxy generation is jointly constrained by both the dense reconstruction and the input aerial images. By incorporating image derived structural cues directly into the construction process, the proxy is optimized to align with observed features across views from the outset, rather than relying on post hoc correction that may be infeasible. This shift from a geometry-only to an image-guided paradigm produces proxies that are inherently consistent with the input imagery, enabling more accurate geometry and more coherent texture mapping.
We realize this formulation through a framework that extracts structural priors from images and dense geometry, and enforces their consistency during proxy optimization. The resulting proxies exhibit improved alignment of structural features, reducing artifacts such as edge duplication, texture drift, and cross view inconsistencies. Experiments on diverse architectural scenes demonstrate that our approach produces compact, geometry texture aligned proxies with substantially improved visual fidelity compared to existing methods.
\end{abstract}



\begin{teaserfigure}
\centering
\includegraphics[width=\linewidth]{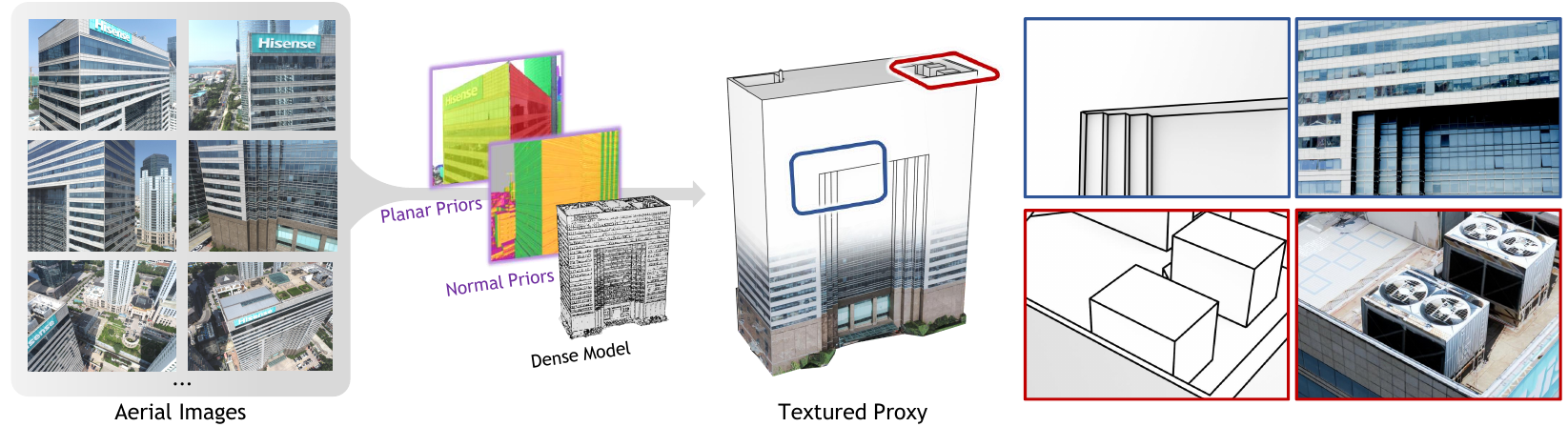}
\caption{Our method generates geometry-texture-aligned architectural proxies from aerial images. While previous approaches generate \textit{Geometry-Only Proxy} deriving solely from the dense reconstruction, our method yields \textit{Image-Guided Proxy} by jointly leveraging the dense model and the aerial images. }
\label{fig:teaser}
\end{teaserfigure}

\maketitle
 
\section{Introduction}
\label{sec:intro}

High-quality 3D reconstruction from images has become widely accessible, producing dense geometric models with detailed appearance~\cite{sinha2008interactive,musialski2013survey}. However, such models remain impractical for downstream applications due to their complexity. Proxy representations address this limitation by abstracting geometry into compact, structured models that enable efficient rendering and manipulation. In architectural scenes, proxies are typically composed of piecewise planar elements, capturing the dominant structure of buildings while serving as the support for texture mapping~\cite{garland1997surface,salinas2015structure,gao2022low,fastinstance2024}.

\begin{figure}
\centering
  \includegraphics[width=\linewidth]{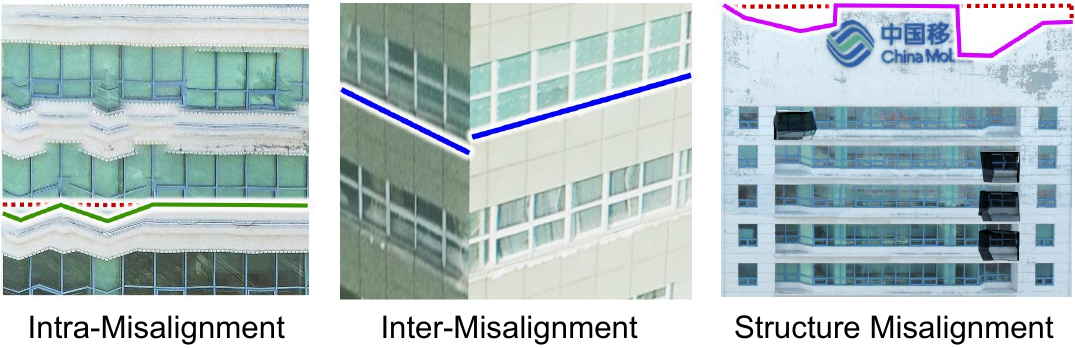}
  \caption{Illustration of three typical issues occurred in \textit{Geometry-Only Proxies} generated with current methods.}
  \label{fig:issues}
\end{figure}

Despite their utility, existing pipelines suffer from a fundamental limitation: geometry abstraction and texturing are treated as separate stages. A proxy is first constructed from a dense model, and textures are later projected from the input images~\cite{TwinTex23,DiffTex25}. This sequential design leads to inconsistencies in what we term \emph{structural features}—planar facades, edges, and their intersections that define the geometry of buildings. These features should coincide across the proxy and all input views, yet in practice they drift. Errors from camera estimation, geometric abstraction, and projection accumulate, causing edges to misalign, planar regions to fragment, and textures from different views to disagree (Fig.~\ref{fig:issues}). As a result, texturing becomes a corrective step that attempts to compensate for misaligned geometry, often with limited success~\cite{TwinTex23,DiffTex25}. Prior approaches either optimize textures without modifying geometry, rely on manual adjustments~\cite{sinha2008interactive}, or employ learning-based generation that may deviate from the observed scene~\cite{chen2025sam}.

We replace this pipeline with a coupled formulation in which proxy geometry is directly constrained by image-derived structural features. 
Rather than constructing a proxy and then fitting textures, we build the proxy so that structural features are already aligned across views, prior to texturing. By enforcing this alignment at construction time, the proxy becomes a geometrically consistent support for all image observations, eliminating the need for downstream correction. 
To realize this formulation, we extract priors from the input images and use them to jointly guide camera refinement and proxy generation. 
Camera poses are optimized to align projected geometry with image-derived normal cues, while proxy geometry is iteratively adjusted under a guidance field that enforces piecewise planar structure consistent with these priors. This shared structural signal couples geometry and appearance throughout the pipeline. Given the resulting aligned proxy and refined cameras, texture mapping reduces to aggregating consistent observations rather than resolving inconsistencies.

Our approach produces textured proxies with significantly improved alignment of structural features, reducing artifacts such as ghosting, edge duplication, and texture drift. We demonstrate consistent gains in both geometric accuracy and visual fidelity over prior methods that decouple geometry and texturing. More broadly, our results suggest that enforcing structural alignment at the level of representation, rather than repairing it afterward, is key to reliable proxy reconstruction.
Our contributions are summarized as:
\begin{itemize}
    \item An automatic framework for generating texture-geometry-aligned proxy models given aerial images. Our novel framework can achieve not only high-fidelity textures but also high-quality proxy geometries;
    \item A building-oriented camera pose refinement method leveraging normal consistency between the input images and reconstructed dense geometries; 
    \item A planar structure segmentation benchmark targeting urban aerial images and a fine-tuned segmentation model. We hope the benchmark can benefit the community;
    \item The first proxy generation method targeting texture-geometry alignment, from a textureless dense building reconstruction. 
\end{itemize}

\section{Related Works}
\label{sec:related-works}

\paragraph{Texture Remapping}
Existing texture remapping methods for reconstructed dense geometry can be classified into two categories: 
i) Image-only remapping methods associate each face or vertex with one or multiple appropriate input images, then create textures via seam/weights optimization to avoid visible seams between adjacent patches~\cite{wang2018seamless,waechter2014let,fu2018texture,bernardini2001high,lee2020texturefusion,bi2017patch},
ii) Joint optimization methods jointly rectify the camera poses and geometric errors, to avoid texture misalignment caused by geometric error and camera drifting~\cite{lempitsky2007seamless,gal2010seamless, zhou2014color,fu2020joint}.

In recent years, texturing architectural proxy models has emerged as a topic of growing interest.
\citet{TwinTex23,DiffTex25} generate a texture map for each planar polygon of an architectural proxy, minimizing the intra-misalignment via image warping or weights optimization.
However, these methods focus only on texture improvement, which subjectively decouples the inter-relationship between camera pose, geometry and texture. These image-only methods still struggle in tackling inter-misalignment and structure misalignment. 

In this paper, we focus on generating a texture-geometry aligned proxy model given aerial images. To the best of our knowledge, we are the first textured proxy generation method that jointly optimizes camera poses, proxy model, and texture maps.

\paragraph{Camera Calibration}
Camera calibration methods can be categorized by optimization granularity. 
Pixel-based methods~\cite{schonberger2016structure,triggs1999bundle} minimize per-pixel reprojection errors and achieve high accuracy, but are sensitive to textureless regions and image noise. 
Primitive-based methods~\cite{hu2024pas,shu2023structure} exploit geometric primitives such as feature points, lines, and planes to formulate more robust alignment objectives, yet their effectiveness depends on reliable primitive detection and association.
In large-scale urban reconstruction, building facades are dominated by weak/repetitive textures and convex/concave structures varying with viewing angles, which yield sparse and unreliable feature correspondences. 
Furthermore, current methods minimize average error over all matched feature points rather than enforcing local geometric alignment on individual building structures. As a result, the camera poses may be globally plausible but locally inaccurate for every single building.

Per-image calibration can be further improved using object silhouettes~\cite{lensch2001silhouette} or image matching.
We introduce a building-oriented camera refinement method that incorporates geometric priors estimated by a foundation model.

\begin{figure*}[t]
    \centering
    \includegraphics[width=\linewidth]{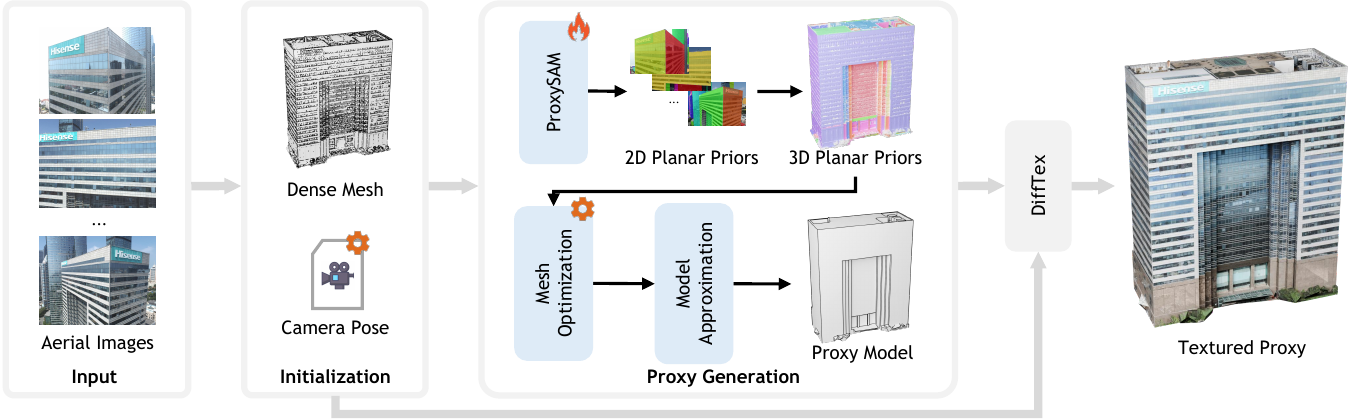}
    \caption{The overview of our method.}
    \label{fig:pipeline}
\end{figure*}

\paragraph{Proxy Model Generation}

Early methods used geometric error metrics or structure-aware decimation techniques~\cite{yan2006quadric, garland1997surface,cohen2004variational,gao2022low,khan2020surface}.
Appearance-preserving simplification approaches seek to approximate the rendered appearance of an existing textured model~\cite{garland1998simplifying, cohen1998appearance,choi2024ltm,liu2025simplifying, bo2026detail}.
Another important family follows a detect-then-assemble paradigm, where planar primitives are first extracted and subsequently assembled into polygonal or polyhedral meshes through optimization~\cite{chauve2010robust,Nan2017polyfit,bauchet2020kinetic, fang2020connect, fang2018planar, kelly2017bigsur}. Representative methods such as PolyFit~\cite{Nan2017polyfit} and KSR~\cite{bauchet2020kinetic} produce compact and structured models, but their performance heavily depends on the completeness and accuracy of primitive detection. Missing or noisy planar hypotheses often lead to structural misalignment. More recently, BLOD~\cite{pan2025building} extends this paradigm to multi-scale LOD generation by introducing an Inside/Outside-view analysis and organizing reconstructed structures into a hierarchical LOD-Tree representation. Despite their effectiveness in geometric abstraction, these approaches largely rely on geometry-only formulations and do not explicitly enforce consistency with image observations during proxy construction.

In contrast to prior geometry-only pipelines, our method incorporates image-derived structural cues directly into proxy optimization. By jointly enforcing geometric abstraction and cross-view image consistency during proxy generation, our approach produces texture-geometry aligned architectural proxies.

\paragraph{Monocular-Plane Extraction}

Estimating planar structures from images has been studied extensively. 
Existing monocular plane extraction methods have achieved notable progress in recovering local planar regions from single images. CNN-based methods~\cite{PlaneNet2018CVPR,PlaneRecover2018ECCV} formulate plane extraction as pixel-wise prediction with a predefined set of planar hypotheses, while instance-level approaches~\cite{PlaneRCNN2019CVPR,PlaneAE2019CVPR} identify multiple plane instances in a scene. More recent Transformer-based and foundation-model-based methods~\cite{PlaneTR2021ICCV,PlaneAC2024,PlaneSeg2024,PlaneSAM2025,ZeroPlane2025} further improve robustness through stronger global reasoning, multiscale context, and large-scale data.
However, these methods are mainly developed for indoor scenes or street-level imagery, where planes are typically observed as local surfaces from close-range viewpoints. In contrast, our goal is building-plane extraction from low-altitude UAV images, which requires globally coherent reasoning over large facade and roof structures under oblique aerial views. The scarcity of fine-grained aerial facade annotations further limits the direct applicability of existing supervised plane extraction pipelines.
\section{Overview}
\label{sec:overview}

Given a set of RGB aerial images $\{ \mathcal{I}_k \}$, we aim at generating an architectural proxy model $\mathcal{M}$ capturing dominant and salient planar structures, along with high-quality texture maps $\mathcal{U}$ aligned with the proxy geometry.

Our framework is illustrated in Fig.~\ref{fig:pipeline}. 
Firstly, we calibrate the images and generate a dense reconstruction. To reduce the geometry-image misalignment, 
the generated camera poses are then refined to minimize the differences between the normal priors estimated from input photos and the normal maps rendered via a differential renderer. 
Next, we construct local/global guidance of dense mesh from planar priors estimated from the input photos via our planar structure segmentation model, called ProxySAM.
Guided by such priors, we optimize the mesh vertices to generate a piece-wise planar building proxy $\mathcal{M}$, via balancing the geometric planarity, preservation of distinct structures, and mesh quality. 
Finally, given the refined camera poses and image-guided proxy, we perform texture remapping to generate the texture maps, $\mathcal{U}$.
\section{Method}

\subsection{Initialization}
\label{sec:pose_opt}

Given the input images $\{\mathcal{I}_k\}$, we first obtain the initial camera pose $\{\mathcal{C}_k\}^0$ and a 3D dense model $\mathcal{D}=(\mathcal{F},\mathcal{V})$ using ContextCapture~\footnote{https://www.soarscape.com/}, following the standard 3D reconstruction pipeline includes: feature mapping, structure from motion (SfM), multi-view stereo (MVS), and surface reconstruction. The generated camera pose $\mathcal{C}_k$ of $\mathcal{I}_k$ consists of a rotation matrix $R_k$ and a translation vector $t_k$. We create a set of binary activation masks $\{ \mathcal{A}_k \}$, each with the same size as $\mathcal{I}_k$, to indicate the target building regions of $\mathcal{D}$. 

\begin{figure}[H]
    \centering
    \includegraphics[width=0.7\linewidth]{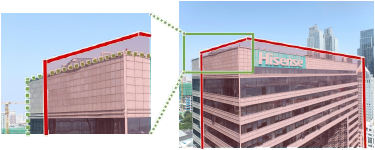}
      \caption{Illusion of building-oriented camera error. The green dash line indicates the building silhouette in the image, while the red region is projection of the reconstructed building model into the image.}
      \label{fig:errors-illustration}
\end{figure}

\paragraph{Building-oriented Camera Error}
Since the calibration method aims to achieve consistency among the observed image features throughout the scene, the initial camera poses $\{\mathcal{C}_k\}^0$ may be globally plausible, but 
locally inaccurate for a single building. Such camera errors bring obvious misalignments between the reconstructed building model $\mathcal{D}$ and the calibrated image views (see the difference between the green image-based building silhouettes and the red building silhouette of reconstructed $\mathcal{D}$ projected to the image in Fig.~\ref{fig:errors-illustration}). The misalignment often led to distinct visual artifacts in the final texture maps and building proxy. 

\paragraph{Normal-based Camera Pose Refinement}
To mitigate such an issue, we propose to refine the camera poses such that each of the input images is geometrically consistent with $\mathcal{D}$. 
As normal maps provide strong geometric cues compared to RGB images~\cite {yu2022monosdf,binninger2025tetweave}, we use the normal maps to guide the refinement of camera poses. 

\begin{figure}
\centering
 \includegraphics[width=\linewidth]{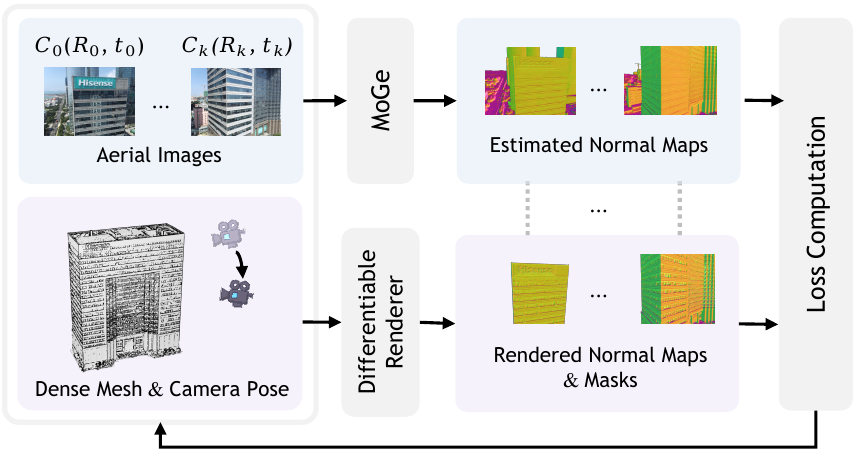}
 \caption{The differential render renders normal maps to compute a loss, which propagates back to refine the camera parameters.}
 \label{fig:pipeline-poseOpt}
\end{figure}

We first estimated the set of normal maps $\{\widehat{\mathcal{N}}_k\}$ of all the input images to serve as normal priors, via a monocular normal estimation model, called MoGe~\cite{wang2025moge}.
Next, for an image $\mathcal{I}_k$ with camera pose $C_k=(R_k, t_k)$, we render $\mathcal{D}$ to produce a normal map $\mathcal{N}_k$ under the identical viewing pose using a differentiable renderer. 
In this way, the differential renderer renders $\mathcal{N}_k$ of each view to compute a loss measuring its distance to the related normal prior $\widehat{\mathcal{N}}_k$, which propagates back to optimize the camera pose $C_k$.
The pipeline of our iterative camera refinement method is illustrated in Fig.~\ref{fig:pipeline-poseOpt}.

\paragraph{Objective Function}
For image $\mathcal{I}_k$, we define the objective function as the weighted sum of a normal loss and a regularization term:
\begin{equation}
L_{Camera} = L_{Normal} + L_{Reg}.
\end{equation}

To minimize the angular discrepancy between the rendered and estimated normals, we introduce the normal loss, $L_{Normal}$, computed as the sum of pixel-wise distances between the rendered and estimated normals over the target building region:
\begin{equation}
L_{Normal} = \frac{1}{|\mathcal{A}_k|}
    \| \mathcal{A}_k \odot ( \widehat{\mathcal{N}}_k - \mathcal{N}_k ) \|_2^2,
\end{equation}
where $\odot$ is the element-wise product.

To ensure numerical stability, we introduce a regularization term \cite{carlone2015initialization} that penalizes excessive deviations of the camera pose from the initial value, $C_k^0=(\mathbf{R}_k^0, \mathbf{t}_k^0)$:
\begin{equation}
 L_{Reg} = \|\mathbf{t}_k - \mathbf{t}_k^0\|_2^2 + \| \log(\mathbf{R}_k^\top \mathbf{R}_k^0)\|_2^2.
\end{equation}

The camera pose of each image is optimized separately. We use Adam~\cite{kingma2014adam} with an initial learning rate of 0.005, running for 1000 iterations per view, and decay the learning rate exponentially every 100 iterations with $\gamma = 0.995$.
\subsection{Proxy Generation}

After obtaining accurate camera poses and dense geometry, our next goal is to transform the dense model $\mathcal{D}=(\mathcal{F},\mathcal{V})$ into a structured proxy $\mathcal{M}$ that aligns with the images. 

Previous approaches derive proxy models solely from the intermediate dense representation (i.e., dense mesh, point clouds)~\cite{Nan2017polyfit, bauchet2020kinetic,pan2025building, cohen2004variational}.
These methods struggle to distinguish structural planes from the noisy dense model, generating inaccurate proxy geometry that heavily deviates from the building shape (see Fig.~\ref{fig:issues}).
We propose to generate proxy models by jointly leveraging the dense reconstruction and the aerial images. 
This new approach enforces the alignment of structural features (planar facades, etc) across views during proxy generation, leading to more accurate and consistent structural features.

Our proxy generation method consists mainly of three stages.
Firstly, we extract 2D planar priors and then compute 3D aerial planar priors accordingly (Sec.~\ref{sec:priors}). 
Next, we compute local and global normal guidance based on planar priors (Sec.~\ref{sec:guidance}). 
The mesh vertices of dense model $\mathcal{D}$ are optimized based on the normal guidance to generate a piece-wise planar mesh $\mathcal{D}^*$ (Sec.~\ref{sec:mesh-optimization}). 
Finally, we extract planar polygons from $\mathcal{D}^*$ to generate a proxy model $\mathcal{P}$ (Sec.~\ref{sec:approximation}).
In this section, we will elaborate on each step in detail.

\subsubsection{Planar Priors Extraction}
\label{sec:priors}

Conventional geometry-only methods are sensitive to local noise, whereas the rich structural cues in the images can compensate for this limitation. 
Hence, we extract 2D planar structure from images as priors, then map them onto $\mathcal{D}$ to generate 3D planar priors. 

\begin{figure}
    \centering
    \includegraphics[width=\linewidth]{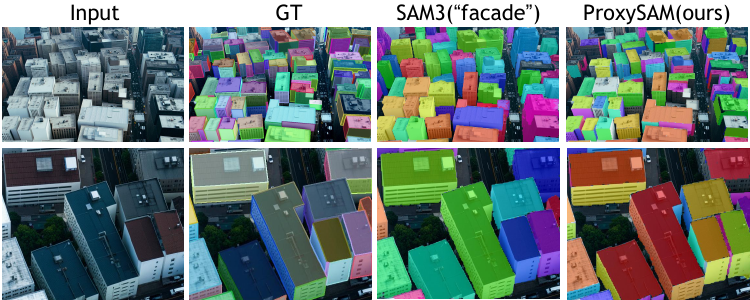}
    \caption{The visualization of two synthetic views and their corresponding facade structure mask in our benchmark. The segmentation results of SAM3 and our ProxySAM.}
    \label{fig:synthetic-plane-seg}
\end{figure}

\paragraph{2D Planar Priors.} 

Existing planar structure segmentation methods and benchmarks target indoor scenes\cite{PlaneNet2018CVPR,PlaneSeg2024,PlaneSAM2025}, or street view images\cite{zhou2020holicity,li2022omnicity}, which failed in aerial images of buildings. 
Successful cases demonstrate the strong semantic understanding ability of Modern fundamental models given sufficient and high-quality training data~\cite{carion2025sam}.
Hence, we constructed \textit{PlaneSet}, a building planar structure benchmark leveraging the Tokyo Dataset~\cite{plateau_tokyo_2023} (covering approximately $3.62\text{ km}^2$). 
We render approximately 1000 aerial views of the Tokyo Dataset using the provided plain proxies.
Next, we apply the region-growing algorithm ~\cite{rabbani2006segmentation} to aggregate faces with similar face normals into the same planar structure. 
After that, we employ Stable Diffusion 3.5~\cite{sd35} conditioned on the rendered plain geometry views via dual ControlNets (Depth and Canny) ~\cite{zhang2023controlnet}, to yield realistic rendering.
Thus, each view has a planar structure mask (see Fig.~\ref{fig:synthetic-plane-seg}).
We then fine-tuned SAM 3~\cite{carion2025sam} using \textit{PlaneSet} via Low-Rank Adaptation (LoRA) \cite{sam3_lora}, to obtain a variant called ProxySAM (see supplementary material for details).

Formally, given an aerial image $\mathcal{I}_k$ and a text prompt “facade plane", ProxySAM predicts corresponding masks of all the distinct planar structures (see Fig.~\ref{fig:synthetic-plane-seg}).
These masks remain fixed in the subsequent stages.

\paragraph{3D Planar Priors.}

For each 2D planar instance mask, we project it onto the dense mesh and obtain a set of triangular faces $\{f\}_m$. We measure the quality, $Q$, of each 2D planar instance mask based on $\{f\}_m$, considering three aspects: face normal consistency, face planarity, and connectivity.
Next, for each frequently observed triangular face, we find the set of instance masks that observe it.
We greedily aggregate adjacent faces into coherent 3D planar regions via region growing~\cite{marshall2001robust, rabbani2006segmentation}. The similarity of two faces is measured by the Jaccard similarity~\cite{manning2008introduction} considering their instance mask set and relative qualities.
Finally, we obtain a partition of $\mathcal{F}$ into planar faces $\{f\}_p$ and non-planar faces $\{f\}_{np}$ 
(see supplementary material for more details).

\subsubsection{Guidance Computation}
\label{sec:guidance}

Next, we compute and assign a guidance normal $G$ to each triangular face that jointly leverages planar priors and geometric cues from the mesh. 
For planar faces, we use global guidance to enforce their smoothness over their planar instance. 
We first estimate the $K$ dominant principal directions by clustering the normals of $\{f\}_p$ using spherical K-means ($K=3$)~\cite{hornik2012spherical}.
These principal directions and their reverse directions together serve as a set of principal normals $\{u\}$. Experimentally, we found that not enforcing orthogonal principal directions yields proxy models that more accurately align with the reconstructed building shapes and texture maps.
For face $f_i \in \{f\}_p$, its guidance normal $G_i$ is initialized as the average normal $\bar{n}$ of its planar instance. We then replace $G_i$ with the closest $u$, if their cosine similarity is larger than $\tau = 0.95$. 
For non-planar faces, we use local guidance to enhance planar structures while preserving sharp structural details.
For non-planar face $f_i \in \{f\}_{np}$, inspired by~\cite{zhang2015guided}, we select the most coherent one-ring patch $p_i$ containing $f_i$. We simplify the patch refinement strategy in \citet{zhong2021shape}, by eliminating the faces whose normal distance from $n_i$ is smaller than a predefined threshold $\tau_n=0.95$ from the patch. 
The normal-wise distance is measured using the dot product.
We then compute the area-weighted average face normal over the refined patch to obtain a representative normal $g_i$ for $f_i$~\cite{zhang2015guided} (see supplementary material for details).
The final local guidance $\mathbf{G}_i$ is computed by aggregating its neighboring normals via joint bilateral weights~\cite{petschnigg2004digital} anchored by representative normals $g$.
Thus, we obtain a per-face guidance normal field $\{G\}$.

\begin{figure}
\centering
    \includegraphics[width=.88\linewidth]{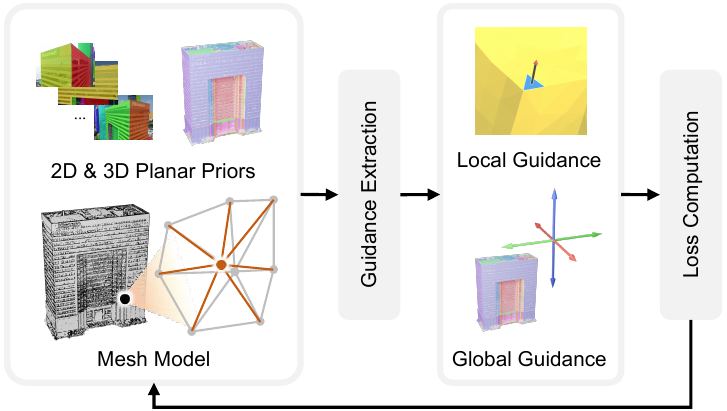}
    \caption{Guided by the 2D and 3D planar priors, the vertices of the dense model are optimized iteratively to generate a piece-wise planar mesh.}
    \label{fig:pipeline-proxy}
\end{figure}

\subsubsection{Normal Guided Model Optimization}
\label{sec:mesh-optimization}

To obtain a piece-wise planar mesh that adheres to the computed guidance field $\{G\}$, we iteratively optimize the vertex positions $V = \{v_j\}$ of $\mathcal{D}$ by minimizing a loss function $L_{Proxy}$:
\begin{equation}
    L_{Proxy} = L_{Planar} + \lambda_l L_{Lap} + \lambda_r L_{PReg},
\end{equation}
where $\lambda_l$ and $\lambda_r$ are the balancing coefficient. We let $\lambda_l = 1$ and $\lambda_r = 100$. The pipeline of optimization is shown in Fig.~\ref{fig:pipeline-proxy}.

We introduce a planar loss term, $L_{Planar}$, to align the face normal $n_i$ of face $f_i$ with its guidance normal $G_i$:
\begin{equation}
    L_{Planar} = \frac{1}{|\mathcal{F}|} \sum_{i} \| n_i - G_i \|_2^2,
\end{equation}
where $|\mathcal{F}|$ denotes the face number of $\mathcal{D}^t$ in iteration $t$.

Meanwhile, we make use of a Laplacian loss term, $L_{Lap}$, to prevent degenerate configurations~\cite{nicolet2021large}:
\begin{align}
    L_{Lap} &= \frac{1}{|\mathcal{V}|} 
    \sum_{v_j \in \mathcal{V}} \bigg\| 
    v_j - \frac{\sum_{m \in H_j }  v_m}{| H_j |}  
    \bigg\|_2^2,
\end{align}
where $|\mathcal{V}|$ denotes the vertex number of $\mathcal{D}^t$ in iteration $t$, $H_j$ denotes the one-ring neighbors of vertex $v_j$, $| H_j |$ denoted the number of adjacent vertices of $v_j$.

\begin{figure}
\centering
 \includegraphics[width=.95\linewidth]{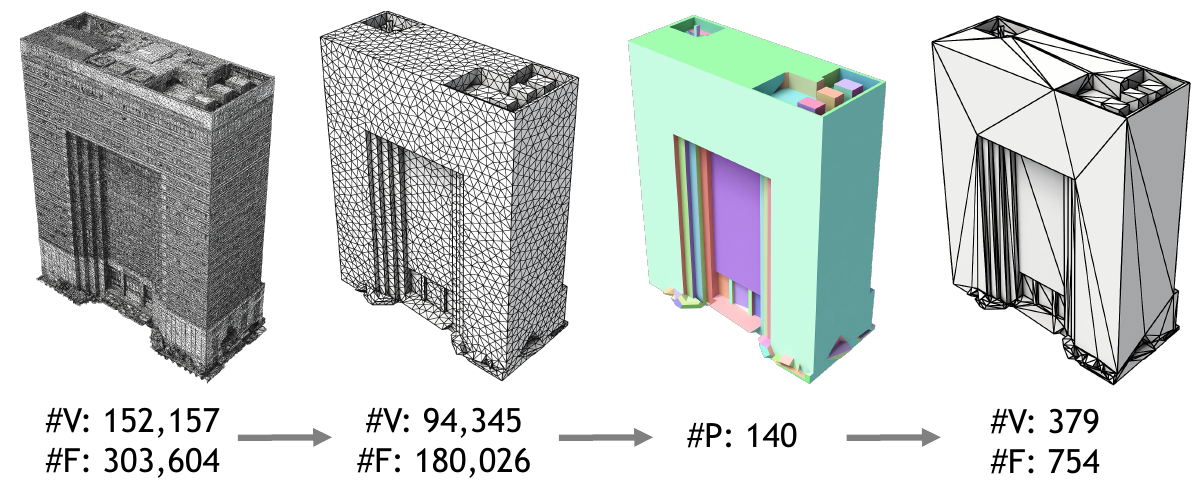}
 \caption{The reconstructed dense mesh is first simplified into a piecewise planar mesh. The planes are organized into proxy polygons and then remeshed into a simplified proxy model. $\# \mathcal{V}$ denotes the number of vertices, $\# \mathcal{F}$ denotes the nunmber of faces, and $\# p$ denotes the number of proxy polygons.}
 \label{fig:mesh-simplificaiton}
\end{figure}

Furthermore, we use a regularization term, $L_{PReg}$, to prevent excessive vertex drift comparing to previous iteration:
\begin{align}  
    L_{PReg} &= \frac{1}{|\mathcal{V}|} 
    \sum_{v_j \in \mathcal{V}} \| v_j^t - v_j^{t-1} \|_2^2,
    \label{eq:loss-laplacian}
\end{align}
where $v^{t-1}_j$ denotes the position of $v_j$ in the previous iteration. 
See supplementary material for the ablation study on $L_{PReg}$.

The $L_{Proxy}$ is minimized using Adam~\cite{kingma2014adam}, with an initial learning rate set to $10^{-5}$. We employ exponential decay with $\gamma = 0.995$. The 3D priors and guidance field will be updated after each iteration.
The optimization converges when the number of iterations reaches the maximum value, or the maximum vertex displacement is negligible.

\paragraph{Remeshing.} 

To simplify the dense mesh, we apply adaptive isotropic remeshing~\cite{pietroni2009almost} every 10 iterations, while setting the target edge length to 110\% of the average edge length of the current mesh. 

\subsubsection{Model Approximation} 
\label{sec:approximation}

Finally, we merge adjacent triangles whose normal difference is less than a threshold into a proxy polygon $p$, using the region growing algorithm~\cite{rabbani2006segmentation}. 
The set of proxy polygons $\{p\}$ is subsequently triangulated~\cite{botsch2010polygon} to generate a compact, piecewise-planar proxy mesh model
, $\mathcal{P}$, as shown in Fig.~\ref{fig:mesh-simplificaiton}.
The representation of $\{p\}$ adheres to the existing proxy texturing frameworks. Hence, we can directly send $\mathcal{P}=\{p\}$ to later texturing methods, without further processing.
\subsection{Texture Remapping}
\label{sec:texture_mapping}

Given the refined camera poses and the optimized proxy mesh $\mathcal{P}=\{p\}$, we adopt the plane-based texture remapping method, DiffTex~\cite{DiffTex25}, to generate a texture map for each proxy polygon $p$. 
With our framework, the calibrated photos and proxy model are aligned within the same coordinate system. This allows users to generate a high-quality textured building proxy in a single workflow, without manual alignment between different coordinate systems (such as BLOD~\cite{pan2025building}).
\section{Experimental Results}
\label{sec:results}

\paragraph{Implementation}
We calibrate the input aerial images with the commercial software, called ContextCapture (CC).
The target building is reconstructed as a mesh manifold $\mathcal{D}$ (no self-intersection, etc.).
Our camera refinement is implemented with Nvdiffrast~\cite{laine2020modular}, and proxy generation is implemented with Nvdiffrast and CGAL. All the presented experimental results are obtained on a desktop computer equipped with an Intel i9-14900k processor with 3.0 GHz, 128 GB RAM and an NVIDIA RTX 4090 GPU.

\paragraph{Dataset}

We evaluate the quality of textured proxies using synthetic models and real-world buildings.
We construct a synthetic dataset, containing six models with various layouts and complexity~\cite{xiong2025aerial}. 
The real-world buildings come from public sources~\cite{TwinTex23}.

\subsection{Stage Results}

\begin{figure*}
    \centering   
    \includegraphics[width=\linewidth]{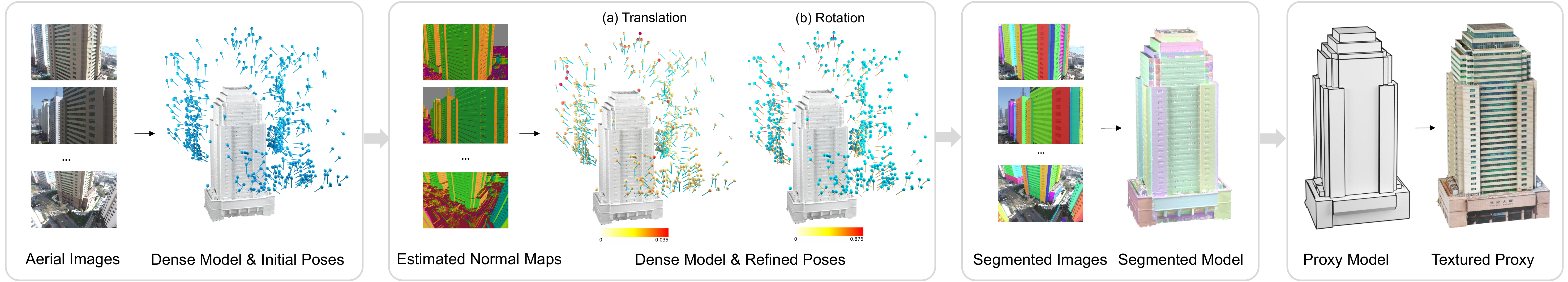}
    \caption{Stage results of our method on the Hall example. From left to right: The input aerial images with estimated normal maps and reconstructed dense model; The refinement of camera poses. The offset of camera poses (left) and viewing angles (right) are colored from small to large (white to red); The generated 2D planar priors (left) and 3D planar priors (right). Each planar instance is assigned a random color; The generated proxy model (left) and textured proxy (right).}
    \label{fig:stage-result}
\end{figure*}

In this section, we provide the stage results of the Hall example in Fig.~\ref{fig:stage-result}. 
Our input is a set of aerial RGB images. We first obtain the initial poses (visualized as a sphere and a cylinder) of aerial images with ContexCapture, reconstruct a dense model, and estimate normal maps by MoGe. We then refine the initial camera pose through translation and rotation. In Fig.~\ref{fig:stage-result}, we visualize the refinement respectively: for translation, the camera spheres are colored by the magnitude of the translation update, while the viewing-direction cylinders are shown in blue; for rotation, the cylinders are colored by the relative rotation angle in degrees between the initial and refined poses, and the camera spheres are shown in blue. Each color map spans the range from zero to the maximum update over all camera poses. Next, the aerial images are segmented by our ProxySAM to obtain 2D planar segmentation. The 3D planar segmentation is then obtained by lifting the 2D predictions through multi-view voting. Finally, we generate a proxy using our image-guided method and obtain the textured proxy based on refined camera parameters.

\begin{table}[t]
\centering
\caption{Quantitative comparison of SSIM and LPIPS metrics. The values are presented as a tuple corresponding to three distinct testing views.}
\label{tab:real_scene_pose_refinement_comparison_quantitative_result-hisense}
\resizebox{\linewidth}{!}{
\begin{tabular}{c c c c c}
\toprule
\multirow{2}{*}{Method} & \multicolumn{2}{c}{SSIM $\uparrow$} & \multicolumn{2}{c}{LPIPS $\downarrow$} \\
\cmidrule(lr){2-3} \cmidrule(lr){4-5}
& Initial Pose & Refined Pose & Initial Pose & Refined Pose \\
\midrule
PolyFit & (0.140, 0.259, 0.211) & (\textbf{0.399}, \textbf{0.433}, \textbf{0.360}) & (0.553, 0.446,0.554) & (\textbf{0.425}, \textbf{0.377}, \textbf{0.439}) \\
KSR & (0.126, 0.286, 0.265) & (\textbf{0.544},\textbf{ 0.534}, \textbf{0.592}) & (0.558, 0.416, 0.518) & (\textbf{0.349}, \textbf{0.334}, \textbf{0.338}) \\
Ours & (0.161, 0.363, 0.264) & (\textbf{0.483}, \textbf{0.422} , \textbf{0.534}) & (0.536, 0.409, 0.513) & (\textbf{0.334},\textbf{0.358}, \textbf{0.362}) \\
\bottomrule
\end{tabular}
}
\end{table}

\subsection{Camera Refinement}

\begin{figure*}
\centering
 \includegraphics[width=\linewidth]{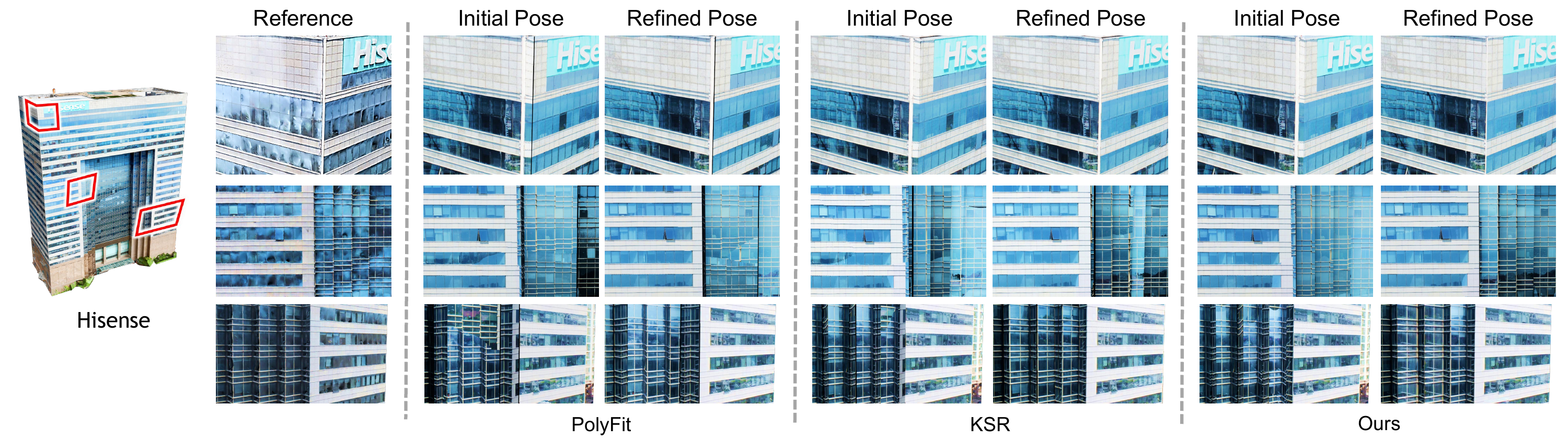}
 \caption{Comparison of the textured proxy models generated with PolyFit, KSR and Ours, given the input aerial images with the initial camera poses and the refined camera poses. The dense reconstruction is visualized as a reference in the first column. The detailed comparisons are shown in the zoomed-in insets.}
 \label{fig:hisense-camera-refinement}
\end{figure*}

Firstly, we evaluate the textured proxy models generated by PolyFit, KSR and our method, given input aerial images with the initial camera poses and refined camera poses of a real scene called Hisense. As shown in Fig.~\ref{fig:hisense-camera-refinement}, the textured proxy given initial poses exhibits distinct texture misalignment. After refinement, such misalignment is substantially reduced for the textured proxy of all methods, validating the effectiveness of our camera refinement.

For the quantitative evaluation of the textured proxy, we adopt the evaluation scheme in~\cite{waechter2014let} to compare rendered images of the reconstructed scenes against the corresponding real images following a specific view from the input cameras. The statistics are provided in Table~\ref{tab:real_scene_pose_refinement_comparison_quantitative_result-hisense}, across all three proxy generation methods, camera pose refinement consistently achieves superior SSIM and LPIPS scores compared to the initial camera baselines. See supplementary material for more comparisons.

\begin{figure}
    \centering
    \includegraphics[width=\linewidth]{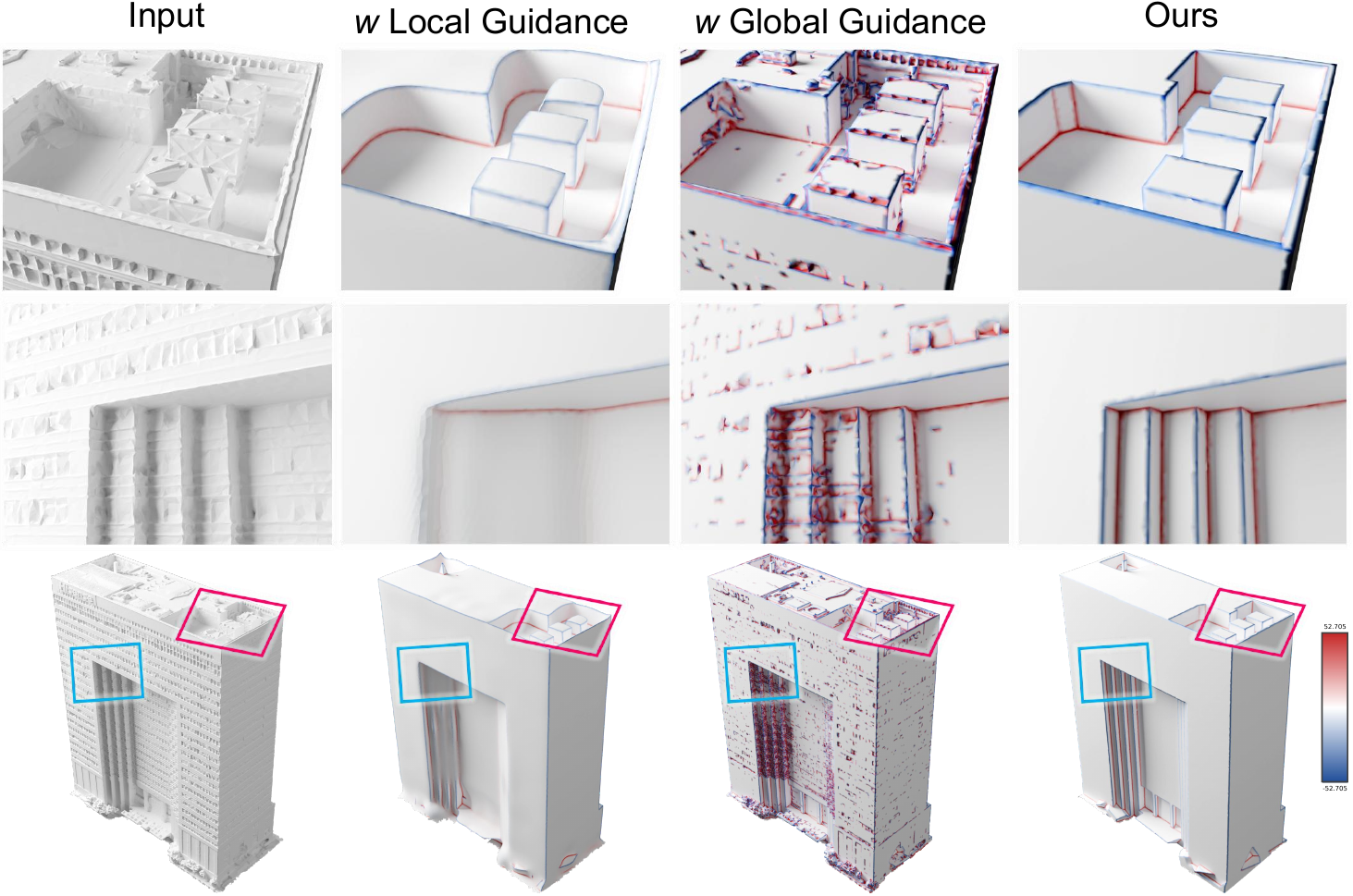}
    \caption{Ablation study on the mesh optimization results regarding the local and global guidance. The colors denote the discrete mean curvature from small to large (blue to red).}
    \label{fig:ablation-proxy}
\end{figure}

\subsection{Proxy Generation}
Next, we estimate the effect of local and global guidance in mesh optimization, as shown in Fig.~\ref{fig:ablation-proxy}.
If we construct the guidance field with only local guidance, the result is overall smooth while preserving sharp edges. However, the local guidance considers only local faces, which cannot distinguish planar structures. As a result, it blends small-scale planar structures and fails to encourage large-scale planarity (second column, Fig.~\ref{fig:ablation-proxy}). 
If we construct the guidance field with only global guidance, dominant planar structures and local structures are preserved, yet small-scale noise is retained (third column, Fig.~\ref{fig:ablation-proxy}). 
After incorporating local and global guidance into the guidance field, the generated mesh model is structurally faithful and piecewise-planar (fourth column, Fig.~\ref{fig:ablation-proxy}).

\subsection{Comparison of Proxy Model}
We compare our method on two challenging synthetic buildings against three proxy generation approaches: VSA~\cite{cohen2004variational} and BLOD~\cite{pan2025building}, which take a dense triangle mesh as input; PolyFit~\cite{Nan2017polyfit}, which takes a point cloud as input. 
For a fair comparison, we tune the parameters of each method to obtain proxies with similar vertex counts. Due to the nature of different algorithms, we cannot guarantee an identical number of vertices.

\begin{figure*}
\centering
  \includegraphics[width=\linewidth]{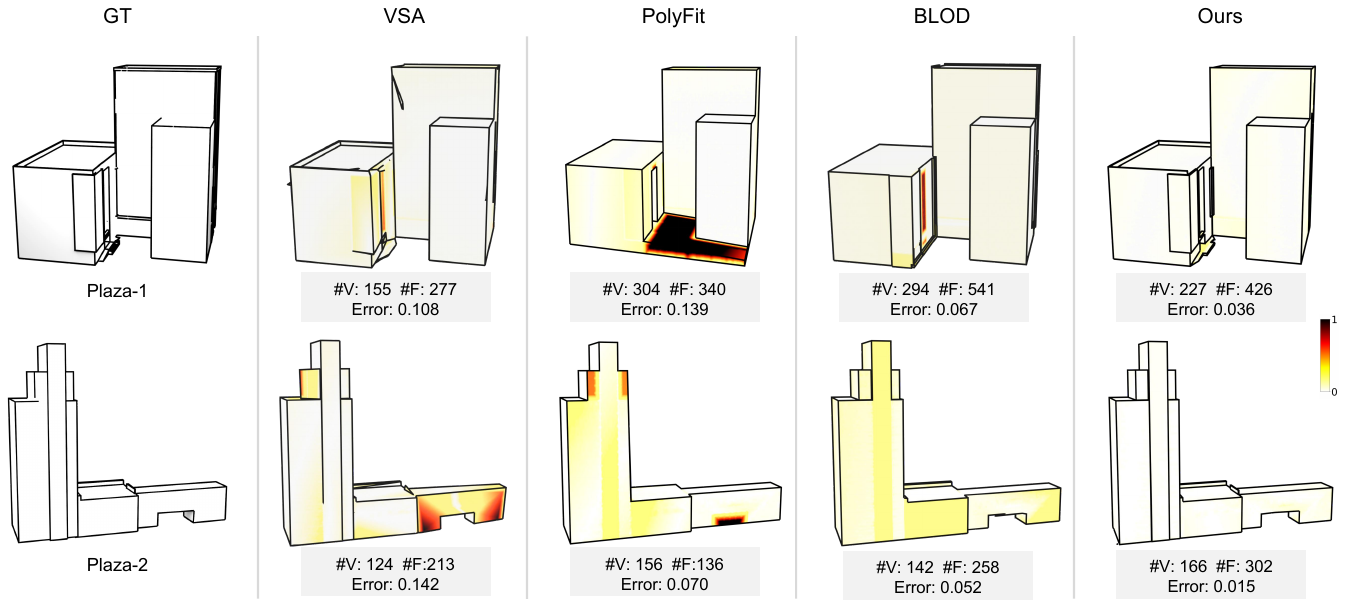}
  \caption{Comparison of the reconstructed proxy models generated with VSA, PolyFit, BLOD and Ours. The ground truth model is visualized as a reference on the left. The colors denote Error values from low (white) to high (dark red).}
  \label{fig:results-virtual-proxy-comparison}
\end{figure*}

Following~\citet{smith2018aerial} and \citet{xiong2025aerial}, we compute \textit{Error}, which quantifies how closely $\mathcal{P}$ approximates the ground-truth model. The results are visualized in Fig.~\ref{fig:results-virtual-proxy-comparison}. 
VSA produces compact approximations through variational surface clustering; however, its region boundaries often exhibit noticeable zigzag artifacts. PolyFit and BLOD recover more regular planar surfaces, but as geometry-only approaches, they rely solely on geometric evidence to determine the planar decomposition. Consequently, stepped or recessed structures with weak geometric cues may be oversimplified, with reconstructed planes spanning across distinct structural levels. 
Our method achieves the lowest Error among all methods, demonstrating the effectiveness of our image-guided method in preserving faithful structural details.

\subsection{Comparison of Textured Proxy}

Serving as geometric support, an accurate proxy model consistent with the input images is critical for the generation of coherent texture mapping. 
Finally, we compare our method with five proxy generation approaches on the textured proxy models: QEM~\cite{garland1997surface}, VSA and BLOD, which take a dense triangle mesh as input; KSR~\cite{bauchet2020kinetic} and  PolyFit, which take a point cloud as input. 
Here, we aim at comparing the visual quality of textured proxies. Instead of simply encouraging similar face/vertex count, for a fair comparison given such noisy dense models as input, we tune the parameters of each method to balance good visual quality with a small face count.
The generated proxies are textured following the identical pipeline given the input aerial images. 
Statistics are provided in Table~\ref{tab:quantitative-results-proxy}. 
See supplementary material for comparison based on similar face/vertex count.

\begin{figure*}
\centering
  \includegraphics[width=\linewidth]{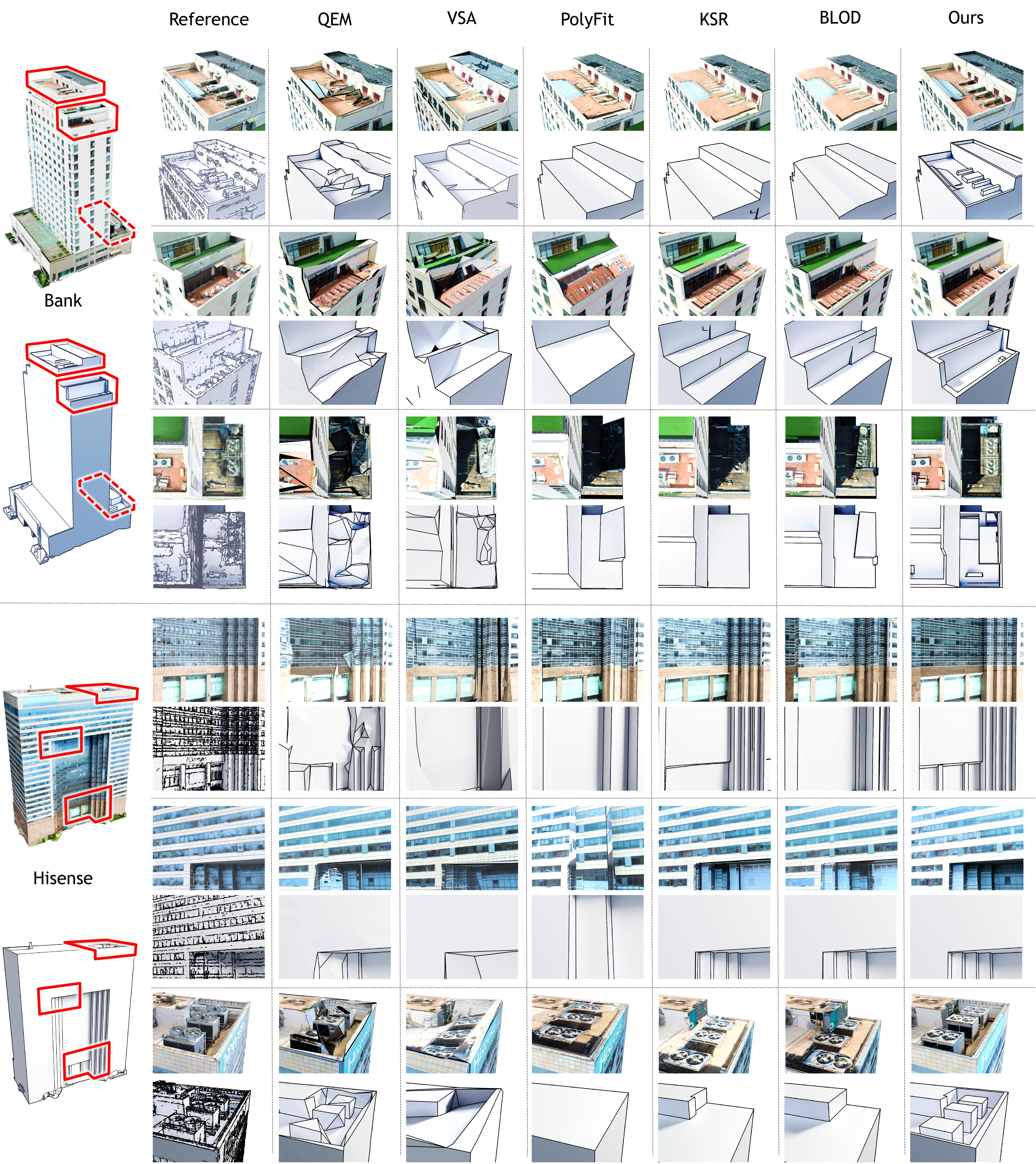}
  \caption{Comparison of the reconstructed proxy models generated with QEM, VSA, PolyFit, KSR, BLOD and ours. The dense reconstruction is visualized as a reference in the first column. The detailed comparisons are shown in the zoomed-in insets.}
  \label{fig:results1}
\end{figure*}

\paragraph{Qualitative Evaluation.}
The qualitative results are visualized in Fig.~\ref{fig:results1}. The reconstructed dense geometry and vertex colors (assigned by CC) are provided as the reference for ground truth structure and colors. 
The QEM progressively simplifies the mesh geometry without considering architectural regularity or planar coherence. 
The generated models exhibit zigzag planar boundaries and irregular structures in all examples.
Meanwhile, VSA approximates the input surface through variational clustering and produces piecewise-smooth regions, but it does not explicitly enforce architectural planarity, leading to locally uneven surfaces on building structures.
In contrast, PolyFit, KSR, and BLOD approximate the dense model using planar structures, generating models consisting of smooth surfaces and sharp edges. 
However, without considering image evidence, they generate fragmented planar patches or spurious planes that span multiple distinct structural surfaces. 
These geometry-only proxy generation methods fail to accurately preserve structural details that align with the ground truth shape (first, fourth, sixth and twelfth rows in Fig.~\ref{fig:results1}).
Among all the results, our image-guided proxies achieve the best balance between simplicity and structural accuracy.

\begin{table*}
    \caption{Quantitative comparison and statistics of the generated textured proxies on real scenes. The top-2 results in proxy quality are highlighted as \colorbox{rankonecolor}{best}, and \colorbox{ranktwocolor}{second}. For the textured proxy quality of each method, the three values in parentheses correspond to the three zoomed-in views in the qualitative results (from top to bottom).}
    \label{tab:quantitative-results-proxy}
    \centering
    {
    \begin{tabular}{c c c c c c c c c c c}
        \midrule[1pt]
        \multirow{2}{*}{Scene} & \multirow{2}{*}{Method} & \multicolumn{3}{c}{Proxy} & \multicolumn{4}{c}{Proxy Quality} & \multicolumn{2}{c}{Textured Proxy Quality}\\
        \cmidrule(lr){3-5} \cmidrule(lr){6-9} \cmidrule(lr){10-11}
        & & $\#\mathcal{V}$  & $\#\mathcal{F}$ &$\# p$  & Error $\downarrow$  & Comp(\%) $\uparrow$ & ECD $\downarrow$ & EF1 $\uparrow$ & SSIM $\uparrow$ & LPIPS $\downarrow$   \\
        \cmidrule(lr){1-2} \cmidrule(lr){3-5} \cmidrule(lr){6-9} \cmidrule(lr){10-11}	
        \multirow{6}{*}{Bank}  
            & QEM & 1026 & 1993 & 427  & \rankone 0.017  & \ranktwo 99.24  & \ranktwo 0.194 & \ranktwo 0.610 & (\,0.645, 0.595, 0.560\,) & (\,0.447, 0.429, 0.503\,) \\
            & VSA & 673 & 1316 & 203  & 0.048  &  96.20 & 0.213 & 0.457 & (0.311, 0.385, 0.300) & (0.493, 0.399, 0.516) \\
            & PolyFit & 400 & 412 & 26  & 0.064  & 88.25  & 0.259 & 0.353 & (\,0.625, 0.567, 0.556\,) & (\,0.573, 0.488, 0.551\,) \\
            & KSR & 767 & 1530 & 221  & 0.044 & 88.12  & 0.210 & 0.408 & (\,0.634, 0.575, 0.597\,) & (\,0.498, 0.495, 0.534\,) \\
            & BLOD & 263 & 505 & 93  & 0.061  & 88.12  & 0.248 & 0.418 & (\,0.638, 0.584, 0.589\,) & (\,0.476, 0.489, 0.521\,) \\
            & Ours & 652 & 1036 & 203  & \ranktwo 0.027  & \rankone 99.88 & \rankone 0.188 & \rankone 0.713 & (\textbf{0.722}, \textbf{0.634}, \textbf{0.608}) & (\textbf{0.311}, \textbf{0.311}, \textbf{0.430}) \\
        \midrule
        \multirow{6}{*}{Hisense} 
            & QEM & 1022 & 1995 & 433  & \rankone 0.016  & \ranktwo 98.40 & \ranktwo 0.056 & \ranktwo 0.506 & (0.557, 0.588, 0.399) & (0.441, 0.534, 0.498) \\
            & VSA & 174 & 327 & 106  & 0.055  & 88.17  & 0.183 & 0.269 & (0.270, 0.233, 0.169)  & (0.495, 0.509, 0.512)  \\
            & PolyFit & 442 & 552 & 14  & 0.106  & 83.90  & 0.072 & 0.385 & (0.557, 0.552, 0.425) & (0.484, 0.623, 0.535) \\
            & KSR & 520 & 1034 & 140 & 0.040  & 89.30 & 0.057 & 0.533 & (0.505, 0.605, 0.464) & (0.561, 0.546, 0.476) \\
            & BLOD & 80 & 156 & 33 &  0.095 & 85.61 & 0.062 & 0.441 & (0.497, 0.596, 0.445) & (0.556, 0.541, 0.508) \\
            & Ours & 379 & 754 & 140 & \ranktwo 0.019  & \rankone 98.90  & \rankone 0.048 & \rankone 0.708 & (\textbf{0.600}, \textbf{0.647}, \textbf{0.529}) & (\textbf{0.402}, \textbf{0.413}, \textbf{0.319}) \\
        \midrule[1pt]
    \end{tabular}
}
\end{table*}%

Moreover, due to the errors that existed in the initial camera poses, there are obvious inter-misalignment (seventh, ninth rows in Fig.~\ref{fig:results1}) and intra-misalignment (seventh and eleventh rows in Fig.~\ref{fig:results1}) in the textured geometry-only proxies, whereas our camera refinement effectively mitigates them. 
Need to mention that texturing the geometry-only proxies with refined camera poses can achieve a significant improvement over both intra-misalignment and inter-misalignment. Please refer to the supplemental material for more results. 

\paragraph{Quantitative Evaluation.}
Following~\citet{smith2018aerial} and \citet{xiong2025aerial}, we compute two metrics to assess the quality of the reconstructed geometry $\mathcal{P}$: i) \textit{Error}, which quantifies how closely $\mathcal{P}$ approximates the dense model $\mathcal{D}$; and ii) \textit{Completeness}, which measures the coverage of $\mathcal{P}$ relative to $\mathcal{D}$. These metrics are computed for all mesh vertices. We further analyze the distributions of these values and report the percentage of vertices with values smaller than 95\% in Table~\ref{tab:quantitative-results-proxy}.
QEM and our method achieve the best or second-best Completeness in all scenes. 
While VSA, PolyFit and BLOD produce a compact proxy, they deliver noticeably higher Error and lower Completeness.
KSR achieves reasonable Error and Completeness, but with a comparable or larger number of faces.

Following~\citet{shen2023flexible} and \citet{chen2022neural}, we use Edge Chamfer Distance (ECD) and Edge F1-score (EF1) to evaluate the preservation of sharp features.
In most cases, our proxies achieve the best ECD and EF1 scores, demonstrating superior preservation of sharp geometric features, such as facade boundaries.

For the evaluation of textured proxies, we compare the rendered images of the reconstructed scenes against the corresponding real images following a specific view from the input cameras~\cite{waechter2014let}.
The comparison results are reported in Table~\ref{tab:quantitative-results-proxy}. 
The rendered views of our textured models achieve the best SSIM and LPIPS scores all the time. 
\section{Conclusion}
\label{sec:conclusion}

Current textured proxy generation methods treat geometry processing and texturing as independent steps, which struggle with texture and geometry misalignment.
In this paper, we propose: i) a framework for generating texture-geometry aligned proxy models given aerial images; ii) A planar structure segmentation benchmark, \textit{PlaneSet}, targeting urban aerial images, and a fine-tuned segmentation model, \textit{ProxySAM}. We hope the benchmark can benefit the community; iii) A building-oriented camera pose refinement method leveraging monocular normal consistency; iv) The first image-guided proxy generation method, targeting texture-geometry alignment.
With our novel framework, we can achieve not only high-fidelity textures but also high-quality geometries.

\paragraph{Limitations and Future Work}

First, our algorithm will preserve unrecognized objects, such as foliage. We plan to incorporate more semantic information to allow the removal of user-specific objects.
Second, we regard the creation of hierarchical planar structures that enable the generation of LODs as future work.
Moreover, unlike performing sequential optimization on geometry and images in our proposed framework, a challenging direction is to jointly optimize the geometry and images for the generation of textured proxy in a single step.

\begin{acks}
This work was supported in part by National Key R\&D Program of China (2024YFB3908500, 2024YFB3908502), NSFC (62302313), ICFCRT (W2441020), Guangdong Basic and Applied Basic Research Foundation (2023B1515120026), Shenzhen Science and Technology Program (KJZD20240903100022028, KQTD20210811090044003), and Scientific Development Fund from Guangdong Provincial Key Laboratory of Visual Media and Multidimensional Intelligence.
\end{acks}

\bibliographystyle{ACM-Reference-Format}
\bibliography{TexProxy}

\end{document}